\documentclass[twocolumn,showpacs,preprintnumbers,prb]{revtex4}

\newcommand{\BE}{\begin{equation}}
\newcommand{\EE}{\end{equation}}
\newcommand{\BA}{\begin{eqnarray}}
\newcommand{\EA}{\end{eqnarray}}
\usepackage{graphicx}
\usepackage{dcolumn}
\usepackage{bm}
\begin{document}

\title{Localization of electromagnetic waves in two-dimensional random dielectric systems}

\author{Bikash C. Gupta} \author{Yu-Yu Chen} \author{Zhen
Ye}\email{zhen@phy.ncu.edu.tw} \affiliation{Wave Phenomena
Laboratory, Department of Physics, National Central University,
Chungli, Taiwan 32054}

\date{January 14, 2003}

\begin{abstract}

We rigorously calculate the propagation and scattering of
electromagnetic waves by rectangular and random arrays of
dielectric cylinders in a uniform medium. For regular arrays, the
band structures are computed and complete bandgaps are discovered.
For random arrays, the phenomenon of wave localization is
investigated and compared in two scenarios: (1) wave propagating
through the array of cylinders; this is the scenario which has
been commonly considered in the literature, and (2) wave
transmitted from a source located inside the ensemble. We show
that within complete band gaps, results from the two scenarios are
similar. Outside the gaps, however, there is a distinct
fundamental difference, that is, waves can be blocked from
propagation by disorders in the first scenario, but such an
inhibition may not lead to inhibition or wave localization in the
second scenario. The study suggests that the traditional method
may be ambiguous in discerning localization effects.

\end{abstract}

 \maketitle

Wave localization is a peculiar property of random media that
completely block wave propagation due to multiple scattering, thus
inducing a surprising phase transition, for example, in optical or
acoustic transparency or electrical conductivity. When localized,
waves remain confined in space until dissipated.

More than two decades have passed since the phenomenon of wave
localization was explored for propagation of electromagnetic (EM)
waves in random media. During this period, a great body of
literature has been generated\cite{reviews}. And the interest in
the subject continues to grow even further in recent
years\cite{weak,Marian,Sigalas,Wiersma,AAA,AAC,debate,Emile}.
Despite the efforts, however, some important problems still remain
unsolved.

The first issue is with the way in which the localization effect
is investigated. To date, claims of localization have been based
on observations of the exponential decay of waves as they
propagate {\it through} disordered media\cite{AAC}. That is, in
most previous experimental or theoretical studies, the apparatus
was set up in such a manner that waves were transmitted at one end
of a scattering ensemble, then the scattered waves were recorded
on the other end to measure the transmission through the sample.
The results were then compared with the theory to infer the
localization effect. In this method, it is quite plausible that
other effects such as reflection and deflection due to the
presence of boundaries may also attenuate waves, resulting in a
similar decay in transmission and obscuring the data
interpretation. Therefore it is desirable to look for a unique
feature which can differentiate localization from other effects;
the inability to discriminate the localization effect from other
effects has caused significant debate in the
literature\cite{debate}.

Second, although it has been suggested a while ago that the
regions of localized states coincide with the positions of the
gaps, the relation between localization in random media and
bandgaps of the corresponding regular systems is still
inconclusive\cite{Sigalas,Datta}.

Third, it has been the prevailing view over the past twenty years
that all EM waves are localized in two dimensions (2D) for any
given amount of disorder, following the scaling analysis of
electronic systems\cite{gang4}. Recently, there is an intensive
debate on this view from new experiments\cite{Pudalov,Shashkin2},
as reviewed in \cite{EA}. Since localization in electronic and EM
systems has the same physical origin, it is therefore imperative
to re-look at the view that all EM waves are always localized in
2D random systems. This task may be difficult, due to the obvious
limitation of the finite sample size for either numerical or
experimental workers, but at least one may examine whether the
phenomenon of localization has been explored in a proper way in
the past.

With this Letter, we wish to shed new light on these questions.
Here, we present a rigorous study of EM wave scattering and
propagation in media containing many dielectric cylinders. The
approach is based upon the self-consistent theory of multiple
scattering\cite{Twersky} and has been used previously to study
acoustic localization in liquid media\cite{Emile} and acoustic
attenuation by rigid cylinders in air\cite{Chen}. Wave propagation
is expressed by a set of coupled exact equations and is solved
rigorously. We show that wave localization can be achieved in
ranges of frequencies, coincident with yet wider than the complete
bandgap. For the phenomenon of wave localization, we compare two
scenarios by analogy with the acoustic case\cite{Scaling3}: (1)
the traditional setup of probing localization both numerically and
experimnentally, as stated in, e.~g. Ref.~\cite{Sigalas,AAC}, that
is, wave propagating through the array of cylinders, and (2) wave
transmitted from a source located inside the ensemble. We show
that within complete band gaps, results from the two scenarios are
similar, whereas there is a distinct qualitative difference
outside the gap. Moreover, when localized, not only are waves
confined near the transmitting source but a unique collective
phenomenon emerges, illustrated by a phase diagram in analogy to
the acoustic system\cite{Emile}.

The system considered here is similar to what has been presented
in \cite{Sigalas}. Assume that $N$ uniform dielectric cylinders of
radius $a$ are placed in parallel in a uniform medium,
perpendicular to the $x-y$ plane. The arrangement can be either
random or regular. For brevity, we only consider the case of the
E-polarization, i.~e. the E-field parallel to the z-direction. The
qualitative features for both E- and H-polarizations are similar.
The scattering and propagation of EM waves can be solved by using
the exact formulation of Twersky\cite{Twersky}. While the details
can be found in \cite{Chen}, here we brief the main procedures. A
unit pulsating line source transmitting monochromatic waves is
placed at a certain position. The scattered wave from each
cylinder is a response to the total incident wave, which is
composed of the direct contribution from the source and the
multiply scattered waves from each of the other cylinders. The
response function of a single cylinder is readily obtained in the
form of the partial waves by invoking the usual boundary
conditions across the cylinder surface. The total wave ($E$) at
any space point is the sum of the direct wave ($E_0$) from the
transmitting source and the scattered wave from all the cylinders.
The normalized field is defined as $T \equiv E/E_0$; thus the
trivial geometrical spreading effect is eliminated.

\input{epsf}
\begin{center}
\begin{figure}[h]
\epsfxsize=2.25in\epsffile{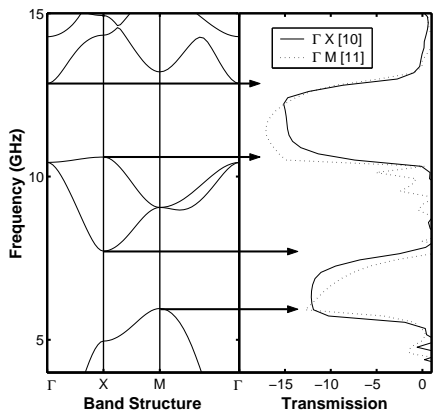}
\caption{Left panel: The band structures computed by the plane
wave expansion method. Right panel: Here is shown the normalized
transmission $\log_{10} |T|^2$ versus frequency; the solid line
refers to the result from the [10] direction propagation, and the
dotted line to that from the [11] direction propagation lines.}
\label{fig1}
\end{figure}
\end{center}

In line with \cite{Sigalas}, the following parameters are used in
the computation. The ratio of the dielectric constant between the
cylinders and the hosting medium is 10; the dielectric constant of
the medium is taken as one. The filling factor $\beta$, the
fraction of area occupied by the cylinders per unit area, is 0.28.
The radius $a$ of the cylinders is 0.38 cm. The lattice constant
$d$ of the corresponding square lattice array of the cylinders is
thus about 1.28 cm ($d = a\sqrt{\pi/\beta}$). For convenience, we
scale all lengths by the lattice constant $d$. The computation is
continued until the convergence is reached.

First, in Fig.~\ref{fig1} we show the band structure of the
corresponding square lattice arrangement of the cylinders,
obtained by the plane wave method. The wave transmission in two
symmetric directions is also shown. Two complete bandgap regions
are identified and are consistent with the highly attenuated
regions in the transmission computation. These results are also
consistent with that in Fig.~4 of \cite{Sigalas}, thereby
verifying our numerical scheme.

\begin{center}
\begin{figure}[hbt]
\epsfxsize=2.25in\epsffile{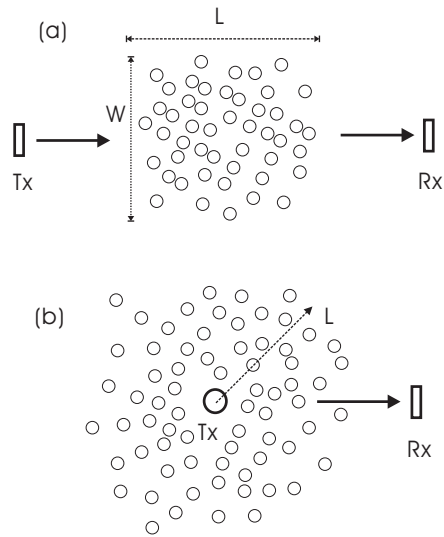} \caption{(a) The `Outside'
case: Electromagnetic propagation through a cloud of dielectric
cylinders. (b) The `Inside' case: Electromagnetic transmission
from a line source located inside the array of dielectric
cylinders.} \label{fig2}
\end{figure}
\end{center}

To investigate wave localization, two situations are considered
and compared: (1) wave propagating through the array of cylinders,
labeled hereafter as the `Outside' situation that imitates the
traditional experimental\cite{Wiersma} and theoretical
setups\cite{Sigalas,AAA}, and (2) wave transmitted from a source
located inside the ensemble, labeled hereafter as the `Inside'
situation. Both cases are illustrated by Fig.~\ref{fig2}. For the
`Outside' case, all cylinders are randomly or regularly placed
within a rectangular area with length $L$ and width $W$. The
transmitter and receiver are located at some distance from the two
opposite sides of the scattering area. For the `Inside' situation,
all cylinders are placed either completely randomly or regularly
within a circle of radius $L$ with the source located at the
center and the receiver located outside the scattering cloud.

\input{epsf}

\begin{center}
\begin{figure}[hbt]
\epsfxsize=2.0in\epsffile{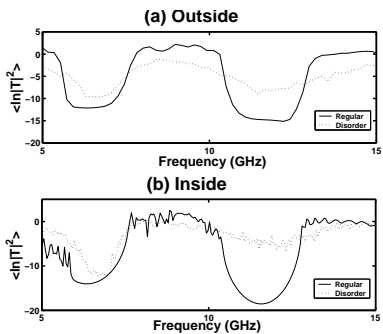} \caption{Transmission
versus frequency for both random and regular arrays of cylinders:
(a) the `Outside' case with W = 6 and L = 10; and (b) the `Inside'
case with L = 10. Please refer to Fig.~\ref{fig2} and the text for
the explanation about the `Outside' and `Inside' cases.}
\label{fig3}
\end{figure}
\end{center}

The frequency response of the averaged logarithmic transmission is
presented in Fig.~\ref{fig3} for both `Inside' and `Outside'
scenarios. Here we see that the disorder somewhat tends to enhance
transmission within the bandgaps for both scenarios, while
obviously reduces the transmission for all frequencies outside the
gaps in the `Outside' case. For the `Inside' situation, however,
the reduction for regions outside the gaps is not generally
obvious, and is only seen near the gap edges. It has been
suggested in the literature that the transmission reduction in the
`Outside' scenario indicates wave localization. For example, the
authors in \cite{Sigalas,AAA} computed the transmission in the
context of the `Outside' scenario, and subsequently obtained the
localization length for all frequencies. We find that this
approach towards localization may not be appropriate. The reasons
follow.

If the transmission reduction in the `Outside' scenario is only
caused by the localization effect, it will be implied that the
random system only supports localized states. Then waves will not
be allowed to propagate not only through but also inside the
system. Therefore we would expect the transmission to follow an
exponential decay with increasing sample size for both `Inside'
and `Outside' setups.

\input{epsf}

\begin{center}
\begin{figure}[hbt]
\epsfxsize=3in\epsffile{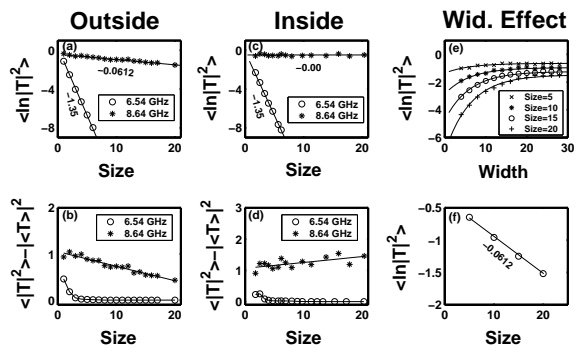} \caption{The averaged
logarithmic transmission and its fluctuation versus the sample
size for two frequencies: one is within the first bandgap and the
other is above the first but below the second gap. The left and
center panels refer to the `Outside' and `Inside' cases
respectively. The estimated slops for the transmission are
indicated in the figure. The right panel shows the effect of width
$W$ and the plot of the transmission versus length $L$ at the
extrapolated infinite width (see the text).} \label{fig4}
\end{figure}
\end{center}

Fig.~\ref{fig4} presents the results for the random ensemble
averaged transmission and its fluctuation as a function of the
sample size at two frequencies. The sample size is varied by
adjusting the number of the cylinders. For the `Outside' case, we
have done the following to remove the effect of the width $W$.
With a fixed sample size (i.~e. the length $L$), we plot the
transmission versus width. We find that the transmission is very
nicely saturating to a certain value in an exponential manner. We
have done for several lengths, and obtained the corresponding
saturated value for each length. Then we plot these values versus
sample lengths. As an example, the results for 8.64 GHz are shown
in Fig.~\ref{fig4} (e) and (f). For 6.54 GHz, the localization is
strong, the width effect diminishes very quickly when the width
increases. Here the plot for 6.54 GHz has width 26 in the
`Outside' scenario. Note that the width should not be started at a
value too close to zero; otherwise the variance will be too large,
making the results unstable. The average has been taken for 500
configuration to ensure the stability.

A few important features are discovered. For the frequency of 6.54
GHz (within the first gap), the transmission decays exponentially
with the sample size for both `Outside' and `Inside' situations
with almost the same slop of -1.35, suggesting that at this
frequency, waves are localized. And inside the localization
regime, the absolute value of the transmission fluctuation is
small, as expected from an earlier work \cite{Emile}. Here we see
that within the localization regime, wave localization can be
indeed observed in both `Outside' and `Inside' scenarios.

For 8.64 GHz (between the first and the second gaps), the
`Outside' and `Inside' scenarios differ significantly. For the
`Outside' case the transmission decreases exponentially with a
slop of -0.0612 along the path. If this exponential decay is
caused by localization, then we should also observe the
exponential decay for the same sample size ($L$) in the `Inside'
scenario. The result in the center panel of Fig.~\ref{fig4}
clearly does not support this point of view. Instead,
Fig.~\ref{fig4} tends to indicate that waves are not yet localized
at 8.64 GHz in the `Inside' scenario. The fact that the
exponential decay only occurs in one scenario but not in the other
for the same sample size ($L$) is itself intriguing and important.
Therefore we may conclude that the `Outside' scenario is
inappropriate in isolating the localization effect, and it would
be a mistake to interpret the exponential decay or transmission
reduction shown in the `Outside' situation as a conclusive
indication of wave localization. Furthermore, as at this
frequency, waves are not yet localized in the `Inside' case and
they have a weaker exponential decay in the `Outside' case, the
transmission will be more sensitive to the arrangement of the
cylinders. Therefore the fluctuation at this frequency is stronger
than that at 6.54 GHz. However, the ratio between the fluctuation
and the transmission at 8.64 GHz can be smaller than that at 6.54
GHz.

\input{epsf}

To this end, a few notes are appropriate. We have also examined
other frequencies in general and two in particular: one is within
the second gap and the other is above the second gap. The results
are very similar to that shown in Fig.~\ref{fig4}. For brevity, we
will not show the results here. From Fig.~\ref{fig3}, the fact
that the transmission reduction occurs not only within but also
outside the gaps (at areas around the edges of the gaps) indicates
that the localized regions are coincident with the complete
bandgaps, and these regions seem wider than the gaps. Our results
show that although the disorders may block waves from propagation
{\it through} the medium, but they may not yet localize the waves
inside a 2D system.

Now we discuss a unique feature of EM wave localization. The
energy flow of EM waves is $\vec{J} \sim \vec{E}\times\vec{H}$. By
invoking the Maxwell equations to relate the electrical and
magnetic fields, we can derive that the time averaged energy flow
is $<\vec{J}>_t \equiv \frac{1}{T}\int_0^T dt \vec{J} \sim
|\vec{E}|^2\nabla\theta,$ where the electrical field is written as
$\vec{E} = \vec{e}_E |\vec{E}|e^{i\theta}$, with $\vec{e}_E$
denoting the direction, $|\vec{E}|$ and $\theta$ being the
amplitude and the phase respectively. It is clear that when
$\theta$ is constant, at least by spatial domains, while
$|\vec{E}| \neq 0$, the flow would come to a stop and the energy
will be localized or stored in the space. We assign a unit phase
vector, $\vec{u} = \cos\theta_i\vec{e}_x + \sin\theta_i\vec{e}_y$
to the oscillation phase $\theta_i$ of the dipoles. Here
$\vec{e}_x$ and $\vec{e}_y$ are unit vectors in the $x$ and $y$
directions respectively. These phase vectors are represented by a
phase diagram in the $x-y$ plane.

\input{epsf}
\begin{center}
\begin{figure}[h]
\epsfxsize=2.25in\epsffile{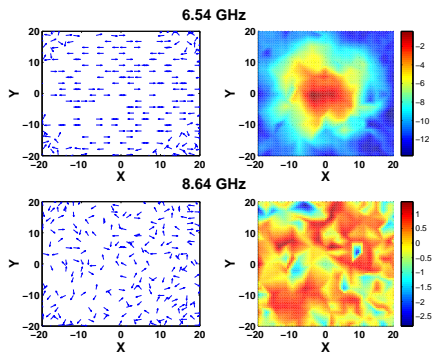} \caption{The phase diagram
and spatial distribution of electromagnetic energy for two
frequencies for one random configuration. Left panel: the phase
diagram for the phase vectors defined in the text; here the phase
of the direct field $E_0$ is set to zero. Right panel: the energy
spatial distribution.} \label{fig5}
\end{figure}
\end{center}

In Fig.~\ref{fig5}, the two-dimensional spatial distribution of EM
energy ($\sim |E/E_0|^2$) and the phase vectors of the E-field are
plotted for the two frequencies discussed in Fig.~\ref{fig4}. The
phase vectors are located randomly in the $x-y$ plane but to avoid
the positions of the cylinders. The `Inside' scenario is
considered. Here we clearly see that for 6.54 GHz, the energy is
mainly confined near the source, consistent with Fig.~\ref{fig4}.
The phase vectors are orderly oriented. These fully comply with
the above general discussion. Therefore at this frequency, EM wave
is indeed localized. When we increasingly add an imaginary part to
the dielectric constant, the ordered orientation of the phase
vectors will disappear, confirming that the phase coherence is a
unique feature of EM wave localization. We note from
Fig.~\ref{fig5} that near the sample boundary, the phase vectors
start to point to different directions. This is because the
numerical simulation is carried out for a finite sample size. For
a finite system, the energy can leak out at the boundary,
resulting in disappearance of the phase coherence. When enlarging
the sample size, we observe that the area showing the perfect
phase coherence will increase. At 8.64 GHz, however, there is no
ordering in the phase vectors $\vec{u}(\vec{r})$. The phase
vectors point to various directions. The energy distribution is
extended in the $x-y$ plane, and no EM wave localization appears,
in agreement with what has been described for Fig.~\ref{fig4}.

In summary, we have examined some fundamental problems of EM wave
localization in 2D. Although it may be still hard to conclude that
extended waves are possible in 2D random media, as limited by the
finite sample size, the present results do indicate that the
traditional method may be unable to isolate the localization
effect. It is also shown that the localization region is related
to and seems to be wider than the complete bandgaps. When
localized, not only are waves confined near the transmitting
source but a unique collective phenomenon emerges.

This work is supported by the National Science Council of Republic
of China.

\end{document}